\documentstyle[preprint,aps]{revtex}
\tightenlines
\begin{document}

\title{Coherence properties of an atom laser}
\author{Marek Trippenbach$^{\,1}$, Y.\ B.\ Band$^{\,1}$,
Mark Edwards$^{\,2}$, Marya Doery$^{\,3}$, and
P.\ S.\ Julienne$^{\,3}$}
\address{$^{\,1}$ Departments of Chemistry and Physics,
Ben-Gurion University of the Negev, Beer-Sheva, Israel 84105\\
$^{\,2}$ Department of Physics, Georgia Southern University,
Statesboro, GA 30458-8031 USA\\
$^{\,3}$ National Institute of Standards and Technology,
100 Bureau Drive Stop 8423, Gaithersburg, MD 20899-8423 USA} 

\date{\today}

\maketitle

\begin{abstract}

We study the coherence properties of an atom laser, which operates by 
extracting atoms from a gaseous Bose-Einstein condensate via a 
two-photon Raman process, by analyzing a recent experiment [(Hagley 
{\it et al.}, submitted to Phys.  Rev.  Lett.  (1999)].  We obtain 
good agreement with the experimental data by solving the 
time-dependent Gross-Pitaevskii equation in three dimensions both 
numerically and with a Thomas-Fermi model.  The coherence 
length is strongly affected by the space-dependent phase developed by 
the condensate when the trapping potential is turned off.
\end{abstract}




One of the most exciting prospects resulting from the Bose-Einstein 
condensation (BEC) of alkali vapors 
\cite{R_JILA_BEC1995,R_MIT_BEC1995,R_RICE_BEC1995} is the possibility 
of producing an intense, coherent, and directed beam of matter waves, 
i.e., an atom laser.  Indeed, prototype atom lasers have already been 
demonstrated \cite{mit_al,nist_al}.  Potential atom laser applications 
include time-and-frequency standards, atom holography, and 
nanolithography.  A critical element in the operation of an atom laser 
is the ``output coupler'' by which atoms are coherently extracted from 
the condensate \cite{Band_oc}.  The design of this element is key to 
controlling the properties of the atom-laser beam \cite{roc_paper}.  
At least two output-coupler mechanisms have been demonstrated.  
Condensate atoms have been extracted by rf magnetic pulses 
\cite{mit_oc,Hansch99} and by two-photon Raman transitions 
\cite{nist_oc}.  A quasi-continuous atom laser was demonstrated 
recently \cite{nist_al} by using a rapid-fire sequence of laser pulses 
each of which caused condensate atoms to undergo a Raman transition 
that transferred momentum while simultaneously changing their internal 
state so that they were not trapped by the magnetic potential.  
Earlier theoretical studies of the properties of atom lasers 
\cite{roc_paper,oc_lit}, made no comparisons of theory and experiment.  
This paper examines the coherence properties of atom laser wavepackets 
by analyzing a recent NIST experiment~\cite{nist_cl_exp} which probes 
such properties by measuring the decay of the interference contrast of 
two overlapping wavepackets outcoupled from a sodium atom condensate 
and separated by a variable delay time $\Delta t$.

A parent condensate with wavefunction $\psi_0({\bf r},t_1)$ is prepared 
at time $t_1$.  In one experiment the harmonic trapping potential was 
left on all the time, and we take $t_1=0$.  In another experiment, the 
trapping potential was turned off and the condensate allowed to expand 
freely for up to $t_1$ = 5 ms.  A 100 ns standing-wave laser pulse was 
applied at time $t_1$ with a wavelength $\lambda_L$ = 589 nm, detuned 
600 MHz to the red of atomic resonance.  This first laser pulse 
diffracts the condensate~\cite{nist_oc} to make two wavepackets 
$\psi_1^\pm$ moving in the $z$ direction with momenta $\pm 2{\bf p}$, 
where $p=\hbar k = h/\lambda_L$.  We consider only the $+2k$ 
wavepacket, since the problem is symmetric.  At time $t_2=t_1+\Delta 
t$, the wavepacket evolves to
\begin{equation}
\psi_1^+({\bf r},t_2) = \phi_1({\bf r}-\Delta{\bf z},t_2) 
e^{i2kz} e^{-i\frac{4E_R}{\hbar}\Delta t} \ ,
\end{equation}
where $E_R=\frac{\hbar^{2}k^2}{2m}$ and $m$ is the atomic mass.  The 
slowly varying envelope function $\phi_1$ is initially just a copy of 
the parent condensate wavefunction with norm $|\alpha|^2 \ll 1$: 
$\phi_1({\bf r},t_1)=\alpha \psi_0({\bf r},t_1)$.  In the experiment 
$|\alpha|^2$ $\approx$ 0.02, and the momentum spread of $\psi_0$ is 
very small compared to $2\hbar k$.  The first wavepacket moves $\Delta 
{\bf z} = {\bf v}\Delta t$ in time $t_2-t_1$, where $v=2\hbar k/m$ is 
the group velocity (60 $\mu$m$/$ms).
 
A second standing wave laser pulse at time $t_2=t_1+\Delta t$ creates 
a second set of wavepackets $\psi_2^\pm$.  The combined number of 
atoms in the $+2k$ wavepacket is 
$\langle|\psi_1^++\psi_2^+|^2\rangle_{\bf r}$, where the brackets 
imply an integration over spatial coordinates.  This fast $+2k$ 
wavepacket soon clears the slowly expanding parent condensate and 
later can be imaged experimentally.  The number of atoms in the $+2k$ 
wavepacket is proportional to the following contrast function 
$C(t_1,\Delta t)$, defined so as to vary between 0 and 1:
\begin{eqnarray}
C(t_1,\Delta t) &=& \frac{1}{4}\left \langle\left|\frac{\phi_1({\bf r}
-\Delta {\bf z},t_2)}{\alpha}e^{-i\frac{4E_R\delta t}{\hbar}}+
\frac{\phi_2({\bf r},t_2)} 
{\alpha}\right|^2\right\rangle_{\bf r} \nonumber \\
&=& \frac{1}{2} + \frac{1}{2}\Gamma(t_1,\Delta t) \ .
\label{eq2}
\end{eqnarray}
The correlation function
\begin{equation}
\Gamma(t_1,\Delta t) ={\rm Re}\left\langle \frac{e^{i\frac{4E_R\delta 
t}{\hbar}}\phi_1^*({\bf r}-\Delta {\bf z},t_2) \phi_2({\bf r},t_2)} 
{|\alpha|^2}\right\rangle_{\bf r}
\label{eq3}
\end{equation}
occuring in Eq.~(2) relates two points separated by distance $\Delta 
{\bf z}$ and provides a measure of the spatial and temporal coherence 
of both the parent condensate and the outcoupled wavepackets.  In the 
hypothetical case that the moving packets are plane waves 
($\frac{\phi_1}{\alpha}=\frac{\phi_2}{\alpha}=1$), then 
$\Gamma(t_1,\Delta t) = \cos(4E_R \Delta t/\hbar)$ varies between $+1$ 
when the wavepackets remain in phase ($4E_R\Delta t/\hbar=2n\pi$ or 
$\Delta t=n\tau$, where $\tau=\frac{h}{4E_R}=10 \, \mu$s for Na atoms) 
and $-1$ when they are out of phase ($\Delta t=(n+\frac{1}{2})\tau$).  
The two packets constructively and destructive interfere in these two 
respective cases, giving a contrast function $C(t_1,\Delta t)$ which 
oscillates between 1 and 0 on a time scale of $\Delta t=\tau/2$.  
Actual condensate wavepackets of finite Thomas-Fermi radius $z_{TF}$ 
in the $z$--direction will physically separate after times on the 
order of $t_{TF}=2z_{TF}/v$, after which $\Gamma\rightarrow 0$ and 
$C(t_1,\Delta t)\rightarrow 1/2$.  Thus $C(t_1,\Delta t)$ oscillates 
rapidly between 1 and 0 when $\Delta t \ll t_{TF}$ and $\Delta z \ll 
z_{TF}$, and approaches $\frac{1}{2}$ when $\Delta t > t_{TF}$ and 
$\Delta z > z_{TF}$.  We will see that when $t_1$ is long enough that 
significant phase modulation has developed across the condensate due 
to the nonlinear mean field, then $C(t_1,\Delta t)$ drops to 
$\frac{1}{2}$ in a time short compared to $t_{TF}$.

In the NIST experiment, the harmonic trap had frequencies 
$\frac{\omega}{2\pi}$ of 14 Hz, $28/\sqrt{2}$ Hz, and 28 Hz in the 
$x$, $y$, and $z$ directions respectively, and a mean frequency of 
$\frac{\bar{\omega}}{2\pi}=28/\sqrt{2}$ Hz.  If the parent condensate 
has 1,500,000 atoms, $z_{TF}(x)= 22$ $\mu$m and $t_{TF}=740$ $\mu$s.  
The characteristic time for developing phase modulation (i.e., 
momentum spread) across the condensate is $1/\bar{\omega}$=8 ms.  The 
experimental $t_1$ varied from 0 to 5 ms, and $\Delta t$ from 0 to 
around 500 $\mu$s.  Wavepacket images were taken about 6 ms after 
$t_1$, long after the fast wavepacket has cleared the stationary, 
slowly expanding parent condensate.  The number of atoms in the $\pm 
2k$ wavepackets could be counted using such images.

In order to provide a reference normalization to remove errors 
incurred by shot-to-shot fluctuations in the total number of 
condensate atoms, since a new condensate had to be made to measure the 
contrast for each delay time, the NIST experiment actually utilized a 
second pair of standing wave pulses to produce a new set of $\pm 2k$ 
wavepackets.  The first pulse of the second pair was applied at time 
$t_3=t_1+3$ms, after the fast wavepackets from the pulse pair at 
$(t_1,t_2)$ have moved away from the parent condensate.  The second 
pulse of the second pair was applied at $t_4=t_3+\Delta t+\tau/2$, 
where $\tau/2=5\mu$s.  Thus, when $\Delta t \ll t_{TF}$, the contrast
function $C_2(t_3,\Delta t+\tau/2)$ for the second pulse pair is 
exactly out of phase with the contrast function $C_1(t_1,\Delta t)$ 
for the first pulse pair.  The experimental images separately 
determine the number of atoms in the $\pm 2k$ wavepackets from the 
$(t_1,t_2)$ and the $(t_3,t_4)$ pulse pairs.  The 
necessity to provide a normalization of the number of atoms each 
packet from shot to shot is avoided in the experiment by reporting as 
the ``signal'' the function
\begin{eqnarray}
S(t_1,t_3,\Delta t) &=& \frac{C_1(t_1,\Delta t)}{C_1(t_1,\Delta t)
+C_2(t_3,\Delta t+\tau/2)} \ .
\end{eqnarray}
Just like $C$, the signal $S$ oscillates rapidly between 0 and 1 for 
$\Delta t \ll t_{TF}$ and approaches $\frac{1}{2}$ when $\Delta t > 
t_{TF}$.  We can define a ``coherence time'' $\Delta t_c$ to be the 
time for the envelope of the signal function $S$ to decay halfway from 
its $\Delta t =0$ value of 1 to its long time limiting value of 
$\frac{1}{2}$, that is, $S(t_1,t_3,\Delta t_c)=0.75$.  A corresponding 
``coherence length'' is $\Delta z_c=v\Delta t_c$.

The time-dependent Gross-Pitaevskii (TDGP) equation describes the 
dynamics of $\psi\left({\bf r},t\right)$, which includes the parent 
condensate plus the fast $\pm 2k$ wavepackets:
\begin{eqnarray}
i\hbar\frac{\partial \psi}{\partial t} &=& 
-\frac{\hbar^{2}}{2m}\nabla^{2}\psi({\bf r},t) +
\left(V_{{\rm trap}}({\bf r},t)+V_{{\rm laser}}({\bf r},t)\right)
\psi({\bf r},t) \nonumber \\
&+& U_{0}N\left|\psi({\bf r},t)\right|^{2} \psi({\bf r},t) \ ,
\end{eqnarray}
where $U_{0}=4\pi\hbar^{2}a/m$, $a$ is the $s$--wave scattering 
length,  $N$ is the total number of condensate atoms and 
$V_{\rm trap}({\bf r},t)$ is the trapping potential. The contrast 
functions $C_1$ and $C_2$ can be calculated from $\psi({\bf r},t)$. 
The interaction of condensate atoms with the four standing-wave laser 
pulses can be written as
\begin{equation}
V_{{\rm laser}}({\bf r},t) = V_{L}
\cos\left(2{\bf k}_{L}\cdot{\bf r}\right)
\sum_{n=1}^{4}
f(t-t_{n},\delta t) \ ,
\label{laser_pulse}
\end{equation}
where $V_{L}f(t,\delta t)$ is the single laser pulse envelope.  The 
laser pulse duration $\delta t = 100$ ns is short compared to $\Delta 
t$.  The factors $t_{n}$ are the times at which the four experimental 
pulses are applied.

The modification to the condensate wavefunction caused by a 
short-time, low-intensity, standing-wave laser pulse can be best 
understood in momentum space.  Before the first pulse, there is only a 
component centered at ${\bf p}=0$.  We can use time-dependent 
perturbation theory to show that sidebands appear after the pulse at 
${\bf p}=\pm 2\hbar{\bf k}$.  For our experimental conditions, it 
can be shown that these sidebands have the same shape as the ${\bf 
p}=0$ component and have amplitude proportional to the pulse area.  At 
short time $\epsilon$ immediately after the $n$-th pulse,
\begin{eqnarray}
\psi({\bf r},t_{n}+\epsilon) 
& \approx &
\psi_{0}({\bf r},t_{n}) - 
\frac{i}{2\hbar} V_{L} {\cal A}(\delta t) \nonumber \\
&&\times
\left[
e^{2i{\bf k}\cdot{\bf r}}+ e^{-2i{\bf k}\cdot{\bf r}}\right]
\psi_{0}({\bf r},t_{n}) \ .
\label{pulse_effect}
\end{eqnarray}
The pulse is applied to the parent condensate at time $t_{n}$, $n$ = 
1,2,3,4, where the parent condensate wavefunction has evolved from 
$t_1$ to $t_n$.  The laser pulse area, $V_{L}{\cal A}(\delta t)$, is 
the area under the single laser pulse appearing in 
Eq.~(\ref{laser_pulse}), and $V_{L}{\cal A}/\hbar \ll 1$ in the 
experiments.  After normalization, the wavepacket in 
Eq.~(\ref{pulse_effect}) serves as the initial condition for 
subsequent evolution.

We have used two different methods to evolve $\psi({\bf r},t)$.  The 
first is a numerical propagation of the three-dimensional (3D) solution to 
the TDGP equation, using a fast-Fourier-transform method with the 
slowly-varying-envelope approximation to reduce the size of the 
spatial grid.  The slowly varying envelope approximation is excellent 
here because the momentum spread of $\psi_0$ is very small compared to 
$2\hbar k$.  We have verified that this methodology, which will be 
described in detail elsewhere, gives excellent agreement with full 
numerical solutions of the TDGP equation in one- and two-dimensions.  
This allows us to calculate exact contrast functions for a zero 
temperature condensate for any time sequence of trapping potential and 
laser pulses.
     
We have also calculated the contrast functions using a second 
approximate method which we call the time-dependent Thomas-Fermi 
(TDTF) method.  Let us first consider the case when the trapping 
potential is turned off at $t=0$ prior to the first pulse at $t_1$.  
Once $V_{\rm trap}$ is removed, the parent condensate, $\psi_{0}({\bf 
r},t)$, evolves freely, develops phase modulation and expands 
somewhat.  The 3D form of $\psi_{0}({\bf r},t)$ can be easily found 
since, for expanding condensates where the TF approximation is valid, 
the solution of the TDGP is self-similar, i.e., it can be transformed 
to its original shape (before release) by suitable axis scalings.  The 
time dependence of the scale parameters has been shown~\cite{CD} to 
obey coupled nonlinear ordinary differential equations.  Once the 
atoms in high momentum states clear the ${\bf p} = 0$ condensate, they 
evolve as free particles (if $V_{L}{\cal A}/\hbar \ll 1$) and move 
with velocity $\pm 2\hbar{\bf k}/m$.  In our 3D model, the full 
condensate wavefunction thus evolves after the passage of the first 
pulse as follows:
\begin{eqnarray}
\psi({\bf r},t) &\approx& \psi_{0}({\bf r},t) - \frac{i}{2\hbar} 
V_{L}{\cal A}(\delta t) e^{-i4E_{R}\left(t - t_{1}\right)/\hbar} 
\nonumber \\
&&
\times \Big[
e^{2i{\bf k}_{L}\cdot{\bf r}}\psi_{0}({\bf r} - {\bf v}(t - t_{1}),t_{1}) 
\nonumber \\ 
&&+\, e^{-2i{\bf k}_{L}\cdot{\bf r}}\psi_{0}({\bf r} + 
{\bf v}(t - t_{1}),t_{1}) \Big].
\label{between_pulses}
\end{eqnarray}
Using Eqs.~(\ref{pulse_effect}) and (\ref{between_pulses}) we can 
develop the condensate wavefunction for any number of pulses and 
delays.  When the trap is left on, the only modification to the above 
analysis is that the ${\bf p} = 0$ condensate does not develop phase 
or expand.

Fig.~\ref{fig1} compares the calculated results for $S(t_1,t_3,\Delta 
t)$ with the NIST data~\cite{nist_cl_exp} for the case where the trap 
was held on.  In each panel the experimental signal is plotted against 
the delay $\Delta t$ used for the first pair of pulses.  The signal 
was measured for $\Delta t$ in 1 $\mu$s increments up to $\Delta t$ = 
50 $\mu$s after which the increment was 30 $\mu$s up to $\Delta t$ = 
530 $\mu$s.  Fig.~\ref{fig1}a shows excellent agreement with the short 
time data, which was normalized to unity at the first peak at 10 
$\mu$s.  The TDGP and TDTF calculations also agree well, except that 
the phase of the latter slightly lags that of the former because of 
the small acceleration of the fast wavepackets by the effective 
potential provided by the parent condensate.  The long-time evolution 
of the signal envelope agrees very well between the TDGP and TDTF 
calculations.  The coherence time $\Delta t_{c}$ predicted by the two 
models, around 275 $\mu$s, is slightly longer than the measured value 
of $225\pm 40$ $\mu$s.  When the trap is on, the decay of 
$S(t_1,\Delta t)$ is simply due to the reduction of the time-dependent 
overlap of the moving outcoupled wavepackets.  Consequently, the 
calculated and measured coherence lengths, $\Delta z_c=v\Delta t_{c}$ 
= 17 $\mu$m and 14 $\mu$m respectively, are an appreciable fraction of 
the size of the parent condensate, $z_{TF}=22$ $\mu$m.  This also 
implies that coherence extends across most of the size of the 
outcoupled wavepackets.  This result is in line with a recent estimate 
of the coherence length of a static condensate using Bragg 
spectroscopy~\cite{mit_bragg}.

Fig.~\ref{fig2} compares the experimental data for $S(t_1,t_3,\Delta 
t)$ with the TDGP and TDTF calculations for two cases for which the 
trap was turned off at $t=0$.  In Fig.~\ref{fig2}a $(t_1,t_3)$ = 
(1.2ms,4.2ms), whereas in Fig.~\ref{fig2}b, $(t_1,t_3)$ = (5ms,8ms).  
The agreement between the two calculations, as well as the agreement 
between experiment and theory, is good for both cases.  The coherence 
times and lengths are much smaller for these trap-off cases than for 
the trap-on case in Fig.~\ref{fig1}.  For Fig.~\ref{fig2}a the 
respective TDGP and TDTF $\Delta t_c$ are 82 and 80 $\mu$s as compared 
to $65 \pm 10$ $\mu$s for the experiment.  For Fig.~\ref{fig2}b the 
corresponding theoretical values of 38 $\mu$s and 37 $\mu$s compare 
with a measured value of $45 \pm 10$ $\mu$s.  The respective coherence 
lengths for the (1.2ms,4.2ms) and (5ms,8ms) cases are 5 $\mu$m and 2 
$\mu$m, much smaller than $z_{TF}$.  Since $\Delta z_c$ is 
substantially smaller than the condensate size, there must be another 
source of coherence loss than wavepacket separation.

The extra source of coherence loss when the trap is off is due to the 
particle interactions that give rise to the nonlinear term in the GP 
equation.  When the trap potential is removed, the parent condensate 
experiences the effective potential $NU_0|\psi_0|^2$, which causes 
phase modulation to develop across the condensate.  This is due to the 
increased spread in the condensate momentum distribution as the atoms 
accelerate.  For example, Fig.~1 of Ref.~\cite{nist_cl_exp} shows the 
spatial oscillations in Re$\psi_0$ and Im$\psi_0$ due to this phase 
modulation.  The presence of these oscillations in $\psi_0({\bf r},t)$ 
spoil the phase matching when packet 1 is translated by $\Delta z$ 
during the interval $\Delta t$, and lead to a much faster loss of 
coherence between the packets than for the trap-on case.  The longer 
$t_1$ is, the greater the coherence loss will be.  Since the 
characteristic time scale to reach terminal momentum spread is 
$1/\bar{\omega}$ = 8 ms, much coherence loss is to be expected for the 
example in Fig.~\ref{fig2}b.

In conclusion, outcoupled wavepacket coherence times and lengths 
predicted by a three-dimensional time-dependent Thomas Fermi model are 
in excellent agreement with those calculated from solution of the 
three-dimensional TDGP equation and also give good 
agreement with data from a recent experiment which measured coherence 
properties of Raman-outcoupled atom laser wavepackets.  Since the 
outcoupled wavepackets are copies of the parent condensate, the 
experiment probes both the coherence of the parent condensate as well 
as that of the outcoupled wavepackets.  Spatial and temporal coherence 
is maintained across the parent condensate while the trap 
is left on, but is rapidly lost when the trap is turned off due to 
phase modulation which develops across the condensate. 

This work was supported by the US-Israel Binational Science 
Foundation, the James Franck Binational German-Israel Program in 
Laser-Matter Interaction (YBB) and the Office of Naval Research and by 
NSF grant no.\ PHY--9802547.  The authors would like to acknowledge 
stimulating discussions with L.\ Deng, E.W.\ Hagley, C.W.\ Clark, S.L.\ 
Rolston, K. Helmerson, W.D.\ Phillips, and K.\ Burnett.  A special 
thanks is due to the NIST (Gaithersburg) laser cooling group for 
providing data in advance of publication.

\begin{figure}[tbh]
\caption{Comparison of calculated TDGP (solid line) and TDTF (dashed 
line) and experimental (points) signal functions 
$S(t_1=0,t_3=3ms,\Delta t)$ versus $\Delta t$ for the case where the 
trap with 1,500,000 atoms was held on during the laser-pulse firings.  
(a) Comparison during the first 50 $\mu$s where the delay was stepped 
in increments of 1 $\mu$s.  (b) Comparison of the TDGP signal with the 
measured signal envelope over the full delay range to 500 $\mu$s.  The 
TDTF model gives essentially the same envelope.}
\label{fig1}
\end{figure} 

\begin{figure}[tbh]
\caption{Comparison of TDGP (solid line) and TDTF (dashed line) signal 
functions $S(t_1,t_3,\Delta t)$ with the data (points) for the 
case where the trap potential was turned off at $t=0$.  (a) $(t_1,t_3$ 
= (1.2ms, 4.2ms) for 500,000 atoms in the trap.  (b) $(t_1,t_3)$ 
= (5ms, 8ms) for 2,500,000 atoms in the trap.}
\label{fig2}
\end{figure}



\begin{references}

\bibitem{R_JILA_BEC1995} 
M.H. Anderson, J.R. Ensher, M.R. Matthews, 
C.E. Wieman, and E.A. Cornell, 
Science {\bf 269}, 198 (1995).

\bibitem{R_MIT_BEC1995} 
K.B. Davis, M.-O. Mewes, M.R. Andrews, 
N.J. van Druten, D.S. Durfee, D.M. Kurn, and W. Ketterle, 
\prl {\bf 75}, 3969 (1995).

\bibitem{R_RICE_BEC1995} 
C.C. Bradley, C.A. Sackett, and R.G. Hulet,
\prl 78, 985 (1997); see also 
C.C. Bradley, C.A. Sackett, J.J. Tollett, and R.G. Hulet, 
\prl {\bf 75}, 1687 (1995); 

\bibitem{mit_al} 
M.R. Andrews, C.G. Townsend, H.-J. Miesner, D.S. Durfee, 
D.M. Kurn, and W. Ketterle, 
Science {\bf 275}, 637 (1997)

\bibitem{nist_al}
E. W. Hagley, L. Deng, M. Kozuma, J. Wen, K. 
Helmerson, S. L. Rolston, and W. D. Phillips, Science {\bf 283}, 1706 (1999).

\bibitem{Band_oc} Y.B. Band, M. Trippenbach and P.S. Julienne, Phys. Rev. 
{\bf A59}, 3823 (1999); Y. Japha, S. Choi, K. Burnett and Y. B. Band, 
Phys.\ Rev.\ Lett.\ {\bf 82}, 1079 (1999).

\bibitem{roc_paper} M. Edwards, D.A. Griggs, P.L. Holman,
C.W. Clark, S.L. Rolston, and W.D. Phillips, J.\ Phys.\ B:
At.\ Mol.\ and Opt.\ Phys., (in press.)

\bibitem{mit_oc}
M.-O. Mewes, M.R. Andrews, D.M. Kurn, D.S. Durfee, C.G. Townsend, and 
W. Ketterle, \prl {\bf 78}, 582 (1997).

\bibitem{Hansch99} I. Bloch, T. W. H\"{a}nsch, and T. Esslinger, Phys. 
Rev. Lett. {\bf 82}, 3008 (1999).

\bibitem{nist_oc}
M. Kozuma, L. Deng, E. W. Hagley, J. Wen, R. Lutwak, K. Helmerson, S. L. 
Rolston, and W. D. Phillips, Phys.  Rev.  Lett.{\bf 82}, 871 (1999).

\bibitem{oc_lit}
R.J. Ballagh, K. Burnett, and T.F. Scott,
\prl {\bf 78} 1607 (1997);
M. Naraschewski, A. Schenzle, and H. Wallis, Phys. Rev. A {\bf 
56}, 603 (1997);
H. Steck, M. Naraschewski, and H. Wallis, Phys. Rev. Lett. {\bf 
80}, 1 (1998).

\bibitem{nist_cl_exp} 
E.W. Hagley, L. Deng, M. Kozuma, K. Helmerson, S.L. Rolston, 
and W.D. Phillips, accompanying paper.

\bibitem{CD} 
Yu. Kagan, B.V. Svistunov, and G.V. Shlyapnikov,
\pra {\bf 54}, R1753 (1996); 
Y. Castin and R. Dum, 
\prl {\bf 77}, 5315 (1996);
Yu. Kagan, E.L. Surkov, and G.V. Shlyapnikov,
\pra {\bf 55}, R18 (1997).

\bibitem{mit_bragg}
J. Stenger, S. Inouye, A.P. Chikkatur, D.M. Stamper--Kurn, D.E.
Pritchard, and W. Ketterle, (unpublished, cond--mat/9901109.)
 

\end{references}
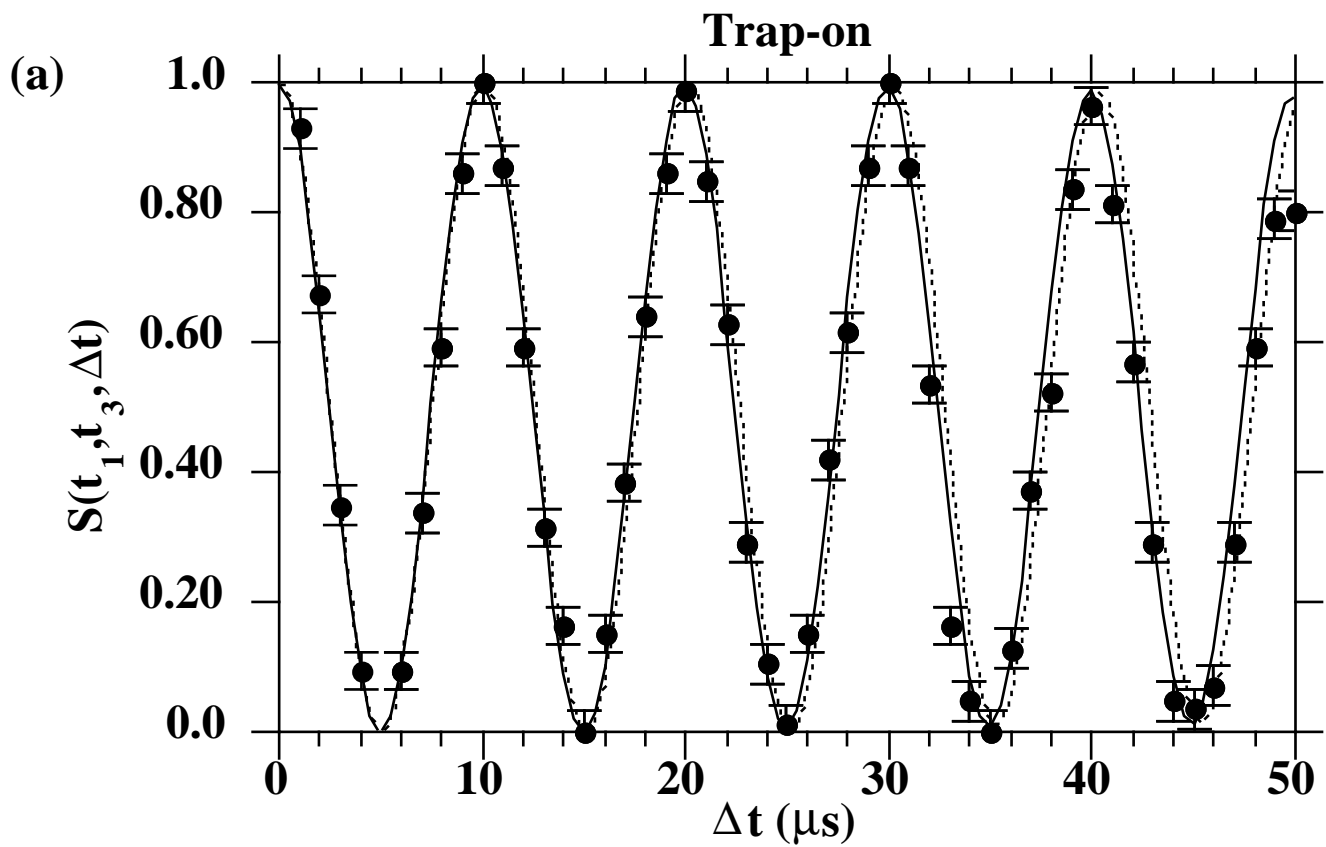
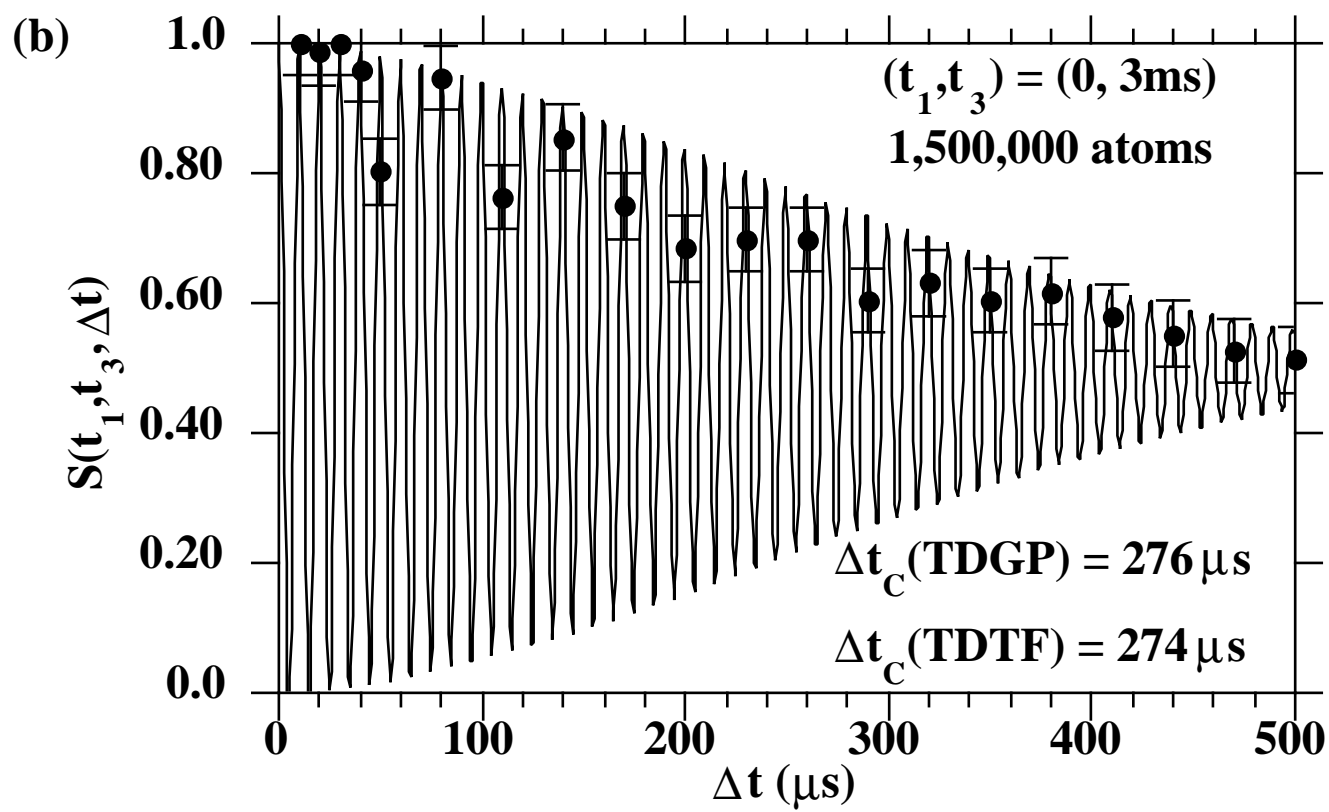
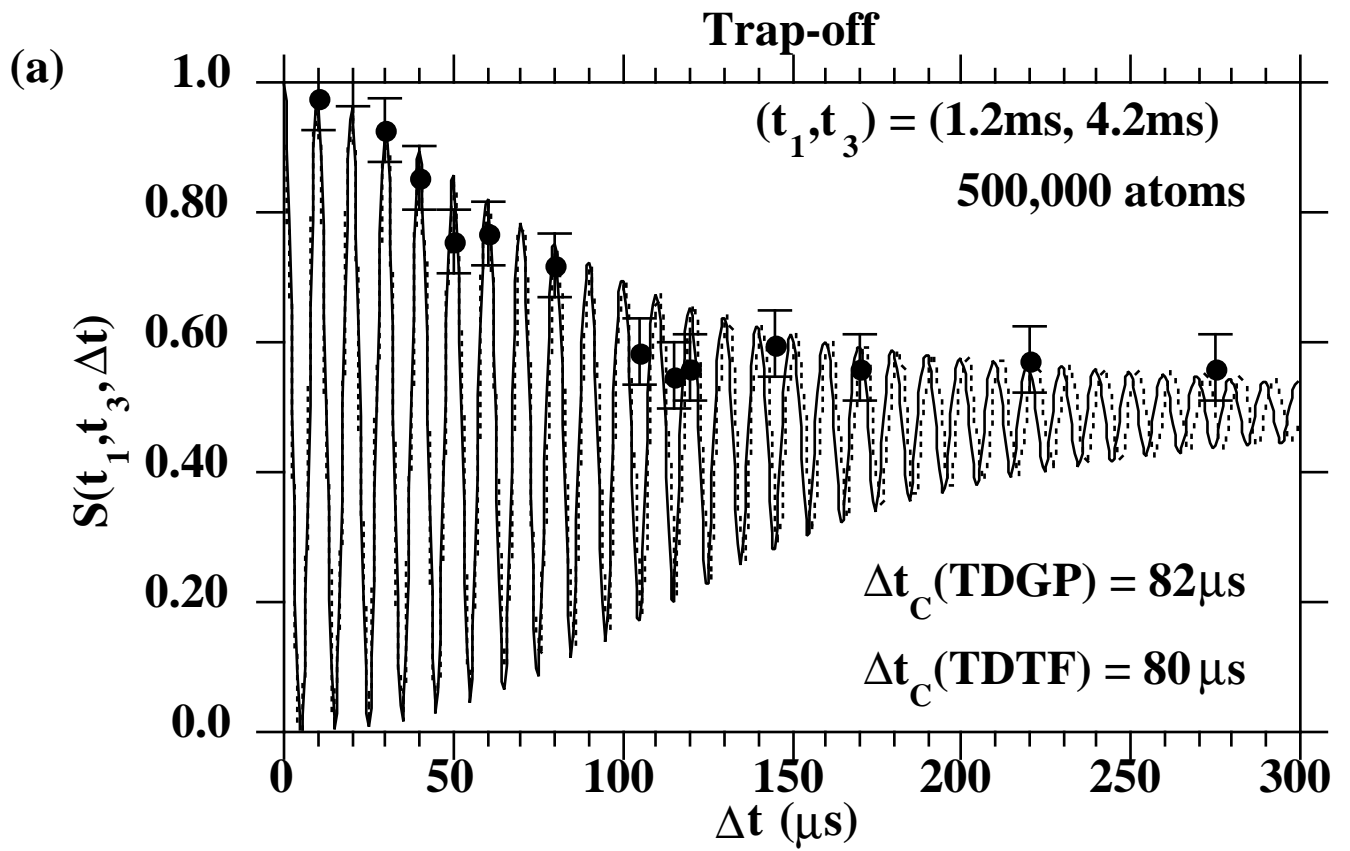
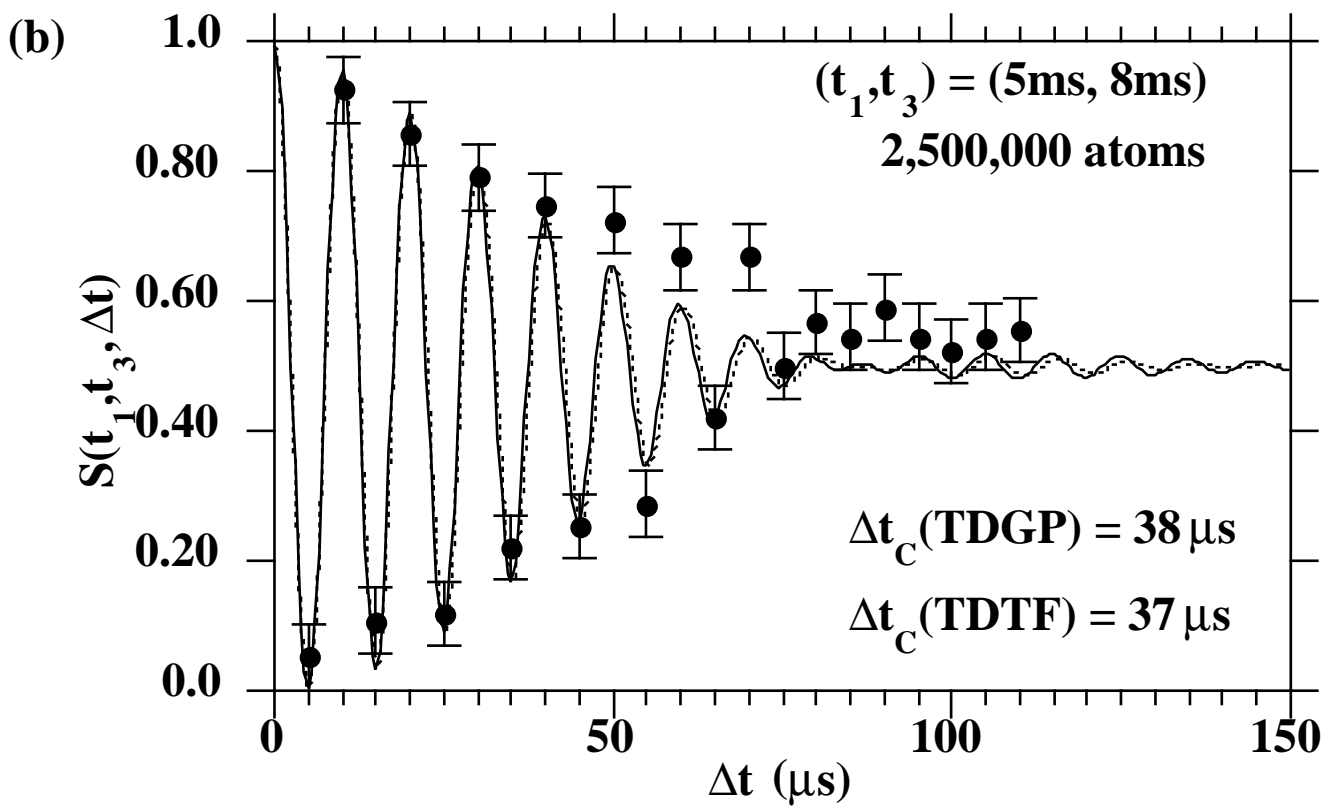
\end{document}